\documentclass[apj]{emulateapj}
\usepackage{apjfonts}
\renewcommand\plotone[1]{\centering\includegraphics[width=\columnwidth]{#1}}

\newcommand\object{\protect}
\submitted{Accepted for publication in ApJ Letters}
\shorttitle{TESTING CRYSTALLIZATION THEORY}
\shortauthors{METCALFE, MONTGOMERY, \& KANAAN}

\begin{document}

\title{Testing White Dwarf Crystallization Theory with Asteroseismology \\
of the Massive Pulsating DA Star BPM~37093}

\author{T. S. Metcalfe\altaffilmark{1}, M. H. Montgomery\altaffilmark{2},
        A. Kanaan\altaffilmark{3}}

\altaffiltext{1}{\footnotesize Harvard-Smithsonian Center for Astrophysics, 
                Cambridge, MA 02138}
\altaffiltext{2}{\footnotesize Institute of Astronomy, University of Cambridge,
                Cambridge, UK}
\altaffiltext{3}{\footnotesize Departamento de F{\'\i}sica, UFSC, 
                Florian{\'o}polis SC, Brazil}

\begin{abstract}

It was predicted more than 40 years ago that the cores of the coolest
white dwarf stars should eventually crystallize. This effect is one of the
largest sources of uncertainty in white dwarf cooling models, which are
now routinely used to estimate the ages of stellar populations in both the
Galactic disk and the halo. We are attempting to minimize this source of
uncertainty by calibrating the models, using observations of pulsating
white dwarfs. In a typical mass white dwarf model, crystallization does
not begin until the surface temperature reaches 6000-8000 K.  In more
massive white dwarf models the effect begins at higher surface
temperatures, where pulsations are observed in the ZZ Ceti (DAV) stars. We
use the observed pulsation periods of \object{BPM~37093}, the most massive
DAV white dwarf presently known, to probe the interior and determine the
size of the crystallized core empirically. Our initial exploration of the
models strongly suggests the presence of a solid core containing about
90\% of the stellar mass, which is consistent with our theoretical
expectations.

\end{abstract}


\section{MOTIVATION}

More than four decades have passed since \cite{abr60}, \cite{kir60} and
\cite{sal61} predicted that the cores of white dwarf stars should
crystallize as they cool down over time. There has never been a direct
empirical test of this theory. The discovery of pulsations in the massive
hydrogen-atmosphere (DA) white dwarf \object{BPM~37093} \citep{kan92}
provided the first opportunity to search for the observational signature
of crystallization in an individual star. Theoretical calculations by
\cite{win97} and \cite{mw99} suggested that the core of this star might be
up to 90\% crystallized, depending on its mass and internal composition.

In addition to providing the first test of the theory of crystallization
in a dense stellar plasma, knowing whether and to what degree this star is
crystallized has implications for a more fundamental question. Recent {\it
Hubble Space Telescope} observations of the faintest white dwarfs in the
globular cluster M4 \citep{han02} have led to a resurgence of interest in
using these stars to provide independent constraints on the ages of
stellar populations. The largest potential sources of error in this method
arise from uncertainties about the composition and structure of white
dwarf interiors. Fortunately, all of the major uncertainties can be
minimized through detailed observation and modeling of pulsating white
dwarfs, providing a crucial method to probe the stellar interiors and
calibrate the cooling models.

The crystallization process leads to one of the largest sources of
uncertainty in the ages of cool white dwarfs \citep{seg94}. When a typical
mass white dwarf star \citep[$0.6~M_\odot$,][]{ngs99} cools down to
$T_{\rm eff}\sim 6000$-8000 K (depending on the
core composition), the high-density core will undergo a phase transition
from liquid to solid. An associated latent heat of crystallization will be
released, providing a new source of thermal energy that introduces a delay
in the gradual cooling of the star \cite[for a recent review,
see][]{hl03}. In mixed C/O cores, phase separation of the ions during
crystallization can provide an additional source of energy, delaying the
cooling even further.

Early attempts to determine the crystallized mass fraction in
\object{BPM~37093} were plagued by difficulties with uniqueness
\citep{mw99}. Recent improvements in our ability to match the observed
pulsation periods in white dwarf stars with theoretical models have been
driven by the development of an optimization method based on a parallel
genetic algorithm \citep{mc03}. This method allows the objective global
exploration of the defining parameters, which is essential to ensure a
unique solution. In this Letter, we present the initial application of
this method to fit the observed pulsation periods of \object{BPM~37093}
with asteroseismological models.

\section{OBSERVATIONS\label{OBS}}

\object{BPM~37093} has been the target of two multi-site observing
campaigns of the Whole Earth Telescope \citep[WET;][]{nat90}. Preliminary
results from these campaigns were published by \cite{kan00}. The 1998
observations ({\sc xcov16}) revealed a set of regularly spaced pulsation
frequencies in the range 1500--2000 $\mu$Hz, which \cite{nit00} identified
as six different radial overtones ($k$), possibly having the same
spherical degree ($\ell=2$). The 1999 observations ({\sc xcov17}) revealed
a total of four independent modes, including two new modes and two which
had been seen in the previous campaign. We obtained new single-site
observations of \object{BPM~37093} from the Magellan 6.5-m telescope on
three nights in February 2003. These data showed evidence of five
independent modes, all of which had been detected in the two previous
multi-site campaigns.

\begin{figure}  
\plotone{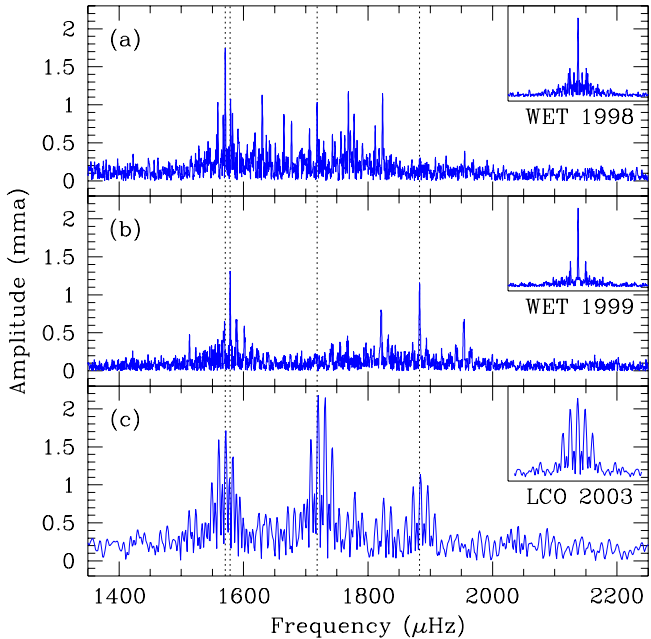}
\caption{Fourier Transforms of the observed light curve of BPM~37093
from multi-site Whole Earth Telescope campaigns in (a) 1998 and (b) 1999,
along with (c) single-site observations from Magellan in 2003. Each   
panel also contains the corresponding window function---the pattern of
alias peaks present in the Fourier Transform for each real frequency.
The amplitudes of various pulsation modes change over time, but their
frequencies are stable.\label{fig1}}
\vspace*{24pt}
\end{figure}


The Fourier Transforms (FTs) of the two WET campaigns and our new
single-site observations are shown in Fig.~\ref{fig1}. Taken together,
these data support the hypothesis---originally proposed by
\cite{kle98}---that each set of observations can provide a subset of the
full spectrum of modes that are actually excited in the star. For reasons
that are not well understood, the amplitudes of many of the pulsation
modes observed in cool DAV stars are highly variable, even though the
frequencies appear to be relatively stable. For example, the three FTs
shown in Fig.~\ref{fig1} all contain the two closely-spaced modes near
1575 $\mu$Hz. However, the mode near 1720 $\mu$Hz in Fig.~\ref{fig1}a was
not detected the following year in Fig.~\ref{fig1}b, but returned with the
same frequency in Fig.~\ref{fig1}c. Likewise, no mode is detected near
1875 $\mu$Hz in Fig.~\ref{fig1}a, while it is clearly present in both
Figs.~\ref{fig1}b and \ref{fig1}c.

We adopt the full set of 8
independent modes proposed by \cite{nit00} for model-fitting. We assume
that each single mode has an azimuthal order $m=0$ \cite[see][]{met03ba}.  
For the three modes consisting of two closely-spaced frequencies, we use
the average of the two. The adopted list of observed periods is shown in
Table \ref{tab1}, along with the periods of two models from our fitting
procedure, which we describe in \S\ref{FIT}.

\section{MODEL FITTING\label{FIT}}

We applied the parallel genetic-algorithm-based fitting method described
in \cite{mc03} to match our white dwarf models to the observed periods of
\object{BPM~37093} listed in Table \ref{tab1}. The extension of this
method from DBV stars to DAV stars primarily involved redefining the
temperature range, and adding an adjustable parameter for the
H layer mass. This led to a significant increase in computing time,
since our code quasi-statically evolves hot polytropic starter models down
to the temperature range of interest---and the DAVs are considerably
cooler ($\sim$10,000 K) than the DBVs ($\sim$25,000 K). To make the
problem computationally tractable, our initial exploration of DAV models
is necessarily limited compared to recent work on DBVs \citep{met03apj}.

\begin{table}
\begin{center}
\tabcaption{Observed and Calculated Periods for BPM~37093\label{tab1}}
\begin{tabular}{lcclcrcclcr}
\tableline\tableline
P$_{\rm obs}$
&~~~~~~~~~~& \multicolumn{4}{c}{1.10 $M_\odot$ (Pure C)}         
&~~~~~~~~~~& \multicolumn{4}{c}{1.10 $M_\odot$ (Pure O)} \\ 
\cline{3-6}\cline{8-11}
      &&$k$ & $\ell$~~~~~ & P$_{\rm calc}$& ~~~~~${\rm O}-{\rm C}$ 
      &&$k$ & $\ell$~~~~~ & P$_{\rm calc}$& ~~~~~${\rm O}-{\rm C}$   \\
\tableline
511.7 && 28 & 2      & 511.58       &$+$0.12 
      && 27 & 2      & 512.50       &$-$0.80 \\
531.1 && 29 & 2      & 530.07       &$+$1.03 
      && 28 & 2      & 530.00       &$+$1.10 \\
548.4 && 30 & 2      & 547.18       &$+$1.22 
      && 29 & 2      & 547.32       &$+$1.08 \\
564.1 && 31 & 2      & 563.71       &$+$0.39 
      && 30 & 2      & 564.89       &$-$0.79 \\
582.0 && 32 & 2      & 581.62       &$+$0.38 
      && 17 & 1      & 582.27       &$-$0.27 \\
600.7 && 33 & 2      & 600.53       &$+$0.17 
      && 32 & 2      & 600.33       &$+$0.37 \\
613.5 && 19 & 1      & 615.23       &$-$1.73 
      && 18 & 1      & 612.61       &$+$0.89 \\
635.1 && 35 & 2      & 636.28       &$-$1.18 
      && 34 & 2      & 636.22       &$-$1.12 \\
\tableline
\end{tabular}
\end{center}
\end{table}

\subsection{Defining the Parameter-Space\label{par}}

The genetic algorithm explored a range of effective temperatures ($T_{\rm
eff}$) between
$
10,000~{\rm K} \le T_{\rm eff} \le 15,000~{\rm K}
$
with a resolution of 100 K. This easily encompasses the empirical DAV
instability strip \citep{ber04}, and allows for any possible shifts in the
temperature scale of our models due to differences in the constitutive
physics \citep[e.g., see][]{mmk03}. Following the recommendation of
\cite{ber95}, we fix the convective efficiency to the ML2/$\alpha$=0.6
prescription.

Recent evolutionary calculations suggest that the He and H
layer masses for massive white dwarfs like \object{BPM~37093} should be
near $\log(M_{\rm He}/M_*)\sim -3.1$ and $\log(M_{\rm H}/M_*)\sim -5.8$
respectively \citep{alt03}. We allowed the genetic algorithm to fit for
He layer masses between
$
-4.0 \le \log(M_{\rm He}/M_*) \le -2.0
$
with a resolution of 0.02 dex. Helium layers with $\log(M_{\rm He}/M_*) >
-2.0$ would theoretically initiate nuclear burning at the base of the
layer.  This would likely result
in models that are too luminous to be consistent with the observations of
\object{BPM~37093}, so we do not consider such thick layers.

A similar consideration leads us to exclude H layer masses with
$\log(M_{\rm H}/M_*)> -4.0$. Our search covers the range
$
-8.0 \le \log(M_{\rm H}/M_*) \le -4.0
$
with a resolution of 0.04 dex, and subject to the constraint that the
H layer be at least two orders of magnitude thinner than the He
layer. This avoids difficulties with overlapping transition zones 
\citep[see][]{bra96}.

Crystallization provides a new energy source for a cooling white
dwarf: the latent heat of crystallization. This supplements
the thermal energy of the ions and delays the cooling, so it primarily
affects the {\it ages} of models at a given $T_{\rm eff}$.
Although the star cools more slowly, the thermal structure at a given
temperature is nearly identical to what it would be in the absence
of crystallization. The change in density due to the
transition from liquid to solid is only a few parts per thousand
\citep{lv75}, so the global mechanical structure of the star is also
relatively unperturbed. While these {\it evolutionary} effects of
crystallization have only a minor influence on the pulsations, the
presence of a solid core can greatly affect the pulsations. The non-radial
$g$-modes are unable to penetrate the solid-liquid interface because the
non-zero shear modulus of the solid effectively excludes their shear
motions, and the oscillations are confined to the fluid regions
\citep{mw99}.

Since we are only interested in fitting the pulsation periods, we can
treat the degree of crystallization as an adjustable parameter for a
given equilibrium model. We simply modify the inner boundary condition
of the pulsational analysis for a given degree of
crystallization. This procedure is an order of magnitude less
computationally expensive than the evolution part of the problem, so we
can afford to repeat it several times. For each evolved model requested by
the genetic algorithm, we calculated the pulsation periods for ten values
of the crystallized mass fraction between
$
0.0 \le M_{\rm cr}/M_* \le 0.9
$
and returned the root-mean-square (rms) differences between the observed
and calculated periods for the value of $M_{\rm cr}$ that minimized the
residuals. The spherical degree ($\ell$) of the modes was also considered
to be unknown, so we calculated all of the $\ell=1$ and $\ell=2$ modes
between 450 and 700~s for each model. Since there are always more
$\ell=2$ modes for a given model, we required them to be closer to the
observed period by a factor $(N_{\ell=2}/N_{\ell=1})$ to be considered a
better match. In effect, we optimized both $M_{\rm cr}$
and the mode identification internally for each model evaluation, while
the genetic algorithm optimized the values of the other three parameters.

In this initial study, we performed fits for three fixed values of the
stellar mass ($M_*$) corresponding to several published estimates.  
\cite{ber95} fit Balmer line profiles to obtain the highest mass estimate
\citep[$M_*=1.10~M_\odot$, according to a recent reanalysis by][]{ber04}.
The lowest mass estimate ($M_*=1.00~M_\odot$) was based on photometric
observations and the trigonometric parallax \citep{blr01}. An intermediate
value ($M_*=1.03~M_\odot$) was obtained by \cite{ka00}, who used a
combination of UV spectra, photometry, and the parallax.

For each of these three masses we performed fits using models with pure C
and pure O cores, which extended to 0.98 $M_r/M_*$ in fractional mass. By
considering only pure core compositions, we do not have to consider the
effect of phase separation of C and O during crystallization, or the
subsequent mixing of the remaining fluid layers \citep[e.g.,
see][]{sal97}. Although these effects could be important, they are
expected to have less impact for large degrees of crystallization, as we
discuss in \S\ref{DISC}.

\subsection{Hare \& Hound Exercises}

Before fitting the observed periods of \object{BPM~37093}, we performed
three `Hare \& Hound' exercises to determine what systematic errors might
arise due to our limited exploration of $M_{\rm cr}$, and from fixing the
mass and the core composition. We constructed a standard $1.00~M_\odot$
C-core model with $T_{\rm eff}=11,700$ K, $\log(M_{\rm He}/M_*)=-3.10$,
$\log(M_{\rm H}/M_*)=-5.76$, and $M_{\rm cr}=0.5~M_*$. We then produced
three new models that each differed from this standard model slightly.
Model `X' had $M_{\rm cr}=0.53~M_*$, model `M' had a
mass of $1.03~M_\odot$, and model `C' had a uniform core of 50\% C and
50\% O. Next, we calculated the periods of each of these modified models
and tried to match them using the genetic algorithm with the mass fixed at
$1.00~M_\odot$ and C-core models.

In each case, the genetic algorithm found an optimal model with parameters
that were shifted slightly from their input values. The optimal match for
model `X' had $M_{\rm cr}=0.5~M_*$, but $T_{\rm eff}$, and the He
and H layer masses were offset by ($-$100 K, $+$0.06 dex, $-$0.04
dex) respectively. This suggests that our coarse sampling of
$M_{\rm cr}$ may lead to small systematic errors in the
other three parameters. The best match for model `C' had the same
$T_{\rm eff}$ as the input model, but the He and H
layer masses and $M_{\rm cr}$ were offset by ($-$0.02
dex, $-$0.32 dex, $-$0.2) respectively. This implies that by performing
fits only with pure C and pure O cores, we may underestimate the actual
value of $M_{\rm cr}$ and the layer masses. The largest offsets came
from the fit to model `M', which overestimated the values by ($+$100 K,
$+$0.72 dex, $+$0.40 dex, $+$0.2) for $T_{\rm eff}$, the He and H
layer masses, and $M_{\rm cr}$ respectively. By fitting
for only a few masses, we might find a spuriously high value of 
$M_{\rm cr}$ and determine the layer masses only to within a factor of 
five.
The above results for models `X' and `M' can be understood in terms of the
calculations of \citeauthor{mw99} (\citeyear[][their \S 7.4 and Eq.~7]{mw99}),
who showed that the average period spacing is much more sensitive to 
changes in $M_*$ than $M_{\rm cr}$.

\begin{table}
\begin{center}
\tabcaption{Fixed-Mass Optimal Models for BPM~37093\label{tab2}}
\begin{tabular}{lrrcrrcrr}
\tableline\tableline
Parameter~~~~~~~~~~~~~~
                  & \multicolumn{2}{c}{1.00 $M_\odot$}            &~~~
                  & \multicolumn{2}{c}{1.03 $M_\odot$}            &~~~
                  & \multicolumn{2}{c}{1.10 $M_\odot$}            \\ 
\cline{2-3}\cline{5-6}\cline{8-9}
                               & Pure C           & Pure O        &
                               & Pure C           & Pure O        &
                               & Pure C           & Pure O        \\
\tableline
$T_{\rm eff}$~(K)$\dotfill$    & 13,700           & 14,500        &
                               & 11,100           & 11,300        & 
                               & 10,500           & 11,500        \\
$\log(M_{\rm He}/M_*)\dotfill$ & $-$2.20          & $-$2.00       &
                               & $-$2.12          & $-$2.26       &
                               & $-$2.60          & $-$2.22       \\
$\log(M_{\rm H}/M_*)\dotfill$  & $-$4.40          & $-$5.00       &
                               & $-$4.28          & $-$4.36       &
                               & $-$4.60          & $-$5.64       \\
$M_{\rm cr}/M_*\dotfill$       & 0.90             & 0.90          &
                               & 0.90             & 0.90          &
                               & 0.90             & 0.90          \\
$\sigma_{\rm P}$~(s)$\dotfill$ & 1.24             & 1.04          &
                               & 1.08             & 1.14          &
                               & 0.95             & 0.86          \\
\tableline
\end{tabular}
\end{center}
\end{table}


\subsection{Application to BPM~37093}

With these limitations firmly in mind, we applied our fitting method to
the observed pulsation periods of \object{BPM~37093}. As mentioned in
\S\ref{par} we repeated the fitting procedure six times: using pure C and
pure O cores for each of three fixed masses. The results of these fits are
shown in Table \ref{tab2}. In every case, the genetic algorithm found a
model that led to rms differences between the observed and calculated
periods ($\sigma_{\rm P}$) near 1~s, which is comparable to the level
of accuracy we achieved with our DBV models \citep{met03apj}.

One of the most striking features of the model-fits in Table \ref{tab2} is
that they {\it all have a crystallized mass fraction of 0.9} within the
resolution of the search ($\Delta M_{\rm cr}=0.1~M_*$). This suggests that
the imprint of a large value of $M_{\rm cr}$ on the pulsation periods
is sufficiently strong to favor a large value of $M_{\rm cr}$ in the fits
{\it regardless of the mass and composition}. The best fits are achieved
for the highest mass models we consider (1.10 $M_\odot$). The calculated
periods and mode identifications for these two models appear with the
observed periods in Table \ref{tab1}. Note that an alternative
identification for the pure O fit to the 582~s period is a ($k=31,
\ell=2$) mode at 582.57~s, degrading the fit to $\sigma_{\rm P}=0.88$~s.

The values of $T_{\rm eff}$ derived from spectral line profile fitting can
provide an independent check of the asteroseismological model-fits listed
in Table \ref{tab2}. The three mass estimates cited in \S\ref{par} also
produced temperature estimates of: $11,550\pm470$ K \citep[for
$1.00~M_\odot$;][]{blr01}, $11,520\pm110$ K \citep[for
$1.03~M_\odot$;][]{ka00}, and $11,730\pm200$ K \citep[for
$1.10~M_\odot$;][]{ber04,fon03}. The temperatures of the lowest
mass model-fits are both significantly (4-6$\sigma$) higher than the
spectroscopic estimate, while the higher mass model-fits are mostly
consistent with the independent measures. Our `Hare \& Hound' exercises
led us to expect some small temperature shifts in the fits, primarily from
fixing the mass and having a limited resolution in $M_{\rm cr}$.

We have no independent method of determining the He and H layer
masses, but the fit values for \object{BPM~37093} fall between the
canonical values of $M_{\rm He}\sim10^{-2}~M_*$ and $M_{\rm
H}\sim10^{-4}~M_*$ \citep{woo92}, and the theoretical values from 
\cite{alt03}.

\section{DISCUSSION \& FUTURE WORK\label{DISC}}

We have conducted the first large-scale exploration of models to fit the
observed pulsation periods of the potentially crystallized white dwarf
\object{BPM~37093}. We have limited this initial study to consider several
fixed values of the stellar mass and the core composition to keep the
problem computationally tractable. Even so, the results shown here
represent more than 6 GHz-CPU-years of calculation, which was only
practical because of the dedicated parallel computers available to this
project \citep{mn00}.

All of our optimal models lead to $M_{\rm cr}=0.9\pm0.1~M_*$. Our `Hare \& 
Hound' exercises suggest that we might be
overestimating $M_{\rm cr}$ for individual fits by fixing the mass, but
also imply that fixing the core composition could cause us to
underestimate $M_{\rm cr}$ by a comparable amount. The best of our six
fits is the 1.10 $M_\odot$ O-core model, which has an rms period
difference of only $\sigma_{\rm P}=0.86$~s relative to the observed
periods of \object{BPM~37093}. The theoretical value of $M_{\rm cr}$
at the temperature, mass, and composition of this model is
$M_{\rm cr}({\rm theor.})=0.93$, suggesting that major revisions to the
existing theory may not be required.

With a significant increase in computing resources, we should be able to
extend this model-fitting method to treat a range of masses spanning the
spectroscopic values. This would have the advantage of eliminating the
potential systematic errors in our fit parameters from fixing the mass.  
As was shown by \cite{mw99}, the periods of the modes are sensitive to
very small changes in $M_{\rm cr}$, suggesting that a
more thorough exploration of this parameter ($\Delta M_{\rm
cr}\sim0.01~M_*$)  would also be fruitful---allowing us to test
crystallization theory with an unprecedented precision.

Realistic stellar models do not predict pure compositions, but rather a
C/O mixture that varies as a function of radius \citep[e.g.,][]{sc93,
sal00}. However, for the large degrees of crystallization which we
have found ($M_{\rm cr}\sim0.9~M_*$), the liquid mantle above the
crystallized core is expected to be fully mixed and significantly enriched
in carbon due to phase separation. As a result, the original C/O profile
in the liquid portion of the core will have been erased, removing any mode
trapping properties which this region would have had on the the pulsation
modes. In this case, the mode trapping would be primarily due to the
envelope transition zones of H and He, simplifying the problem. Thus, in
future fits it might make sense to treat the liquid C/O mantle above the
crystallized core as a uniform mixture of C and O, and to fit for the
relative abundances of these two elements---providing additional insights
about phase separation.

In the near future we expect that the Sloan Digital Sky Survey will
uncover several new massive hydrogen-atmosphere pulsators \citep{muk04},
and each additional object will offer the opportunity to place independent
constraints on crystallization theory. By sampling various masses and
temperatures, we may eventually probe the equations of state of carbon and
oxygen over a broad range of otherwise inaccessible physical conditions.

\acknowledgements

This research was supported by the Smithsonian Institution through
a CfA Postdoctoral Fellowship, by the UK PPARC, and by CNPq through a
PROFIX Fellowship. Computational resources were provided by White
Dwarf Research Corporation through a small grant from NASA 
administered by the American Astronomical Society.


\newpage

\begin{thebibliography}{00000000000000000000000000000000000000000000000000000}

\bibitem[Abrikosov(1960)]{abr60} Abrikosov, A.\ 1960,
Zh.~Eksp.~Teor.~Fiz., 39, 1798

\bibitem[Althaus et al.(2003)]{alt03} Althaus, L.~G. et al. 2003, \aap, 
404, 593

\bibitem[Bergeron et al.(1995)]{ber95} Bergeron, P. et al. 1995, \apj,
449, 258

\bibitem[Bergeron, Leggett, \& Ruiz(2001)]{blr01} Bergeron, P., Leggett,
S.~K., \& Ruiz, M.~T.\ 2001, \apjs, 133, 413

\bibitem[Bergeron et al.(2004)]{ber04} Bergeron, P. et al. 2004, \apj, 
600, 404

\bibitem[Bradley(1996)]{bra96} Bradley, P.~A.\ 1996, \apj, 468, 350

\bibitem[Fontaine et al.(2003)]{fon03} Fontaine, G., Bergeron, P.,
Bill{\`e}res, M., \& Charpinet, S.\ 2003, \apj, 591, 1184

\bibitem[Hansen \& Liebert(2003)]{hl03} Hansen, B.~M.~S.~\& Liebert, J.\
2003, \araa, 41, 465

\bibitem[Hansen et al.(2002)]{han02} Hansen, B.~M.~S.~et al.\ 2002, \apjl,
574, L155

\bibitem[Kanaan et al.(1992)]{kan92} Kanaan, A., Kepler, S.~O.,
Giovannini, O., \& Diaz, M.\ 1992, \apjl, 390, L89

\bibitem[Kanaan et al.(2000)]{kan00} Kanaan, A. et al. 2000, Baltic
Astronomy, 9, 87

\bibitem[Kirzhnitz(1960)]{kir60} Kirzhnitz, D. A.\ 1960, Soviet 
Phys.---JETP, 11, 365

\bibitem[Kleinman et al.(1998)]{kle98} Kleinman, S.~J.~et al.\ 1998, \apj,
495, 424

\bibitem[Koester \& Allard(2000)]{ka00} Koester, D.~\& Allard, N.~F.\
2000, Baltic Astronomy, 9, 119

\bibitem[Lamb \& Van Horn(1975)]{lv75} Lamb, D.~Q.~\& Van Horn, H.~M.\
1975, \apj, 200, 306

\bibitem[Metcalfe(2003a)]{met03ba} Metcalfe, T.~S.\ 2003a, Baltic
Astronomy, 12, 247

\bibitem[Metcalfe(2003b)]{met03apj} Metcalfe, T.~S.\ 2003b, \apjl, 587,
L43

\bibitem[Metcalfe \& Charbonneau(2003)]{mc03} Metcalfe T. S., \&
Charbonneau P. 2003, J.~Comput.~Phys., 185, 176

\bibitem[Metcalfe, Montgomery, \& Kawaler(2003)]{mmk03} Metcalfe, T.~S.,
Montgomery, M.~H., \& Kawaler, S.~D.\ 2003, \mnras, 344, L88

\bibitem[Metcalfe \& Nather(2000)]{mn00} Metcalfe, T.~S.~\& Nather, R.~E.\
2000, Baltic Astronomy, 9, 479

\bibitem[Montgomery \& Winget(1999)]{mw99} Montgomery, M. H. \& Winget, 
D. E. 1999, \apj, 526, 976

\bibitem[Mukadam et al.(2004)]{muk04} Mukadam, A. et al. 2004, \apj,
submitted

\bibitem[Napiwotzki, Green, \& Saffer(1999)]{ngs99} Napiwotzki, R., Green,
P. J.  \& Saffer, R. A. 1999, \apj, 517, 399

\bibitem[Nather et al.(1990)]{nat90} Nather, R. E. et al. 1990, \apj, 361, 
309

\bibitem[Nitta(2000)]{nit00} Nitta, A. 2000, Ph.D.~Thesis, The University 
of Texas at Austin

\bibitem[Salaris et al.(2000)]{sal00} Salaris, M. et al. 2000, \apj, 544, 
1036

\bibitem[Salaris et al.(1997)]{sal97} Salaris, M. et al. 1997, \apj,
486, 413

\bibitem[Salpeter(1961)]{sal61} Salpeter, E.~E.\ 1961, \apj, 134, 669 

\bibitem[Segretain \& Chabrier(1993)]{sc93} Segretain, L.~\& Chabrier, G.
1993, \aap, 271, L13

\bibitem[Segretain et al.(1994)]{seg94} Segretain, L. et al. 1994, \apj,
434, 641

\bibitem[Winget et al.(1997)]{win97} Winget, D.~E. et al. 1997, \apjl, 
487, L191

\bibitem[Wood(1992)]{woo92} Wood, M.~A.\ 1992, \apj, 386, 539

\end{thebibliography}
\end{document}